\documentclass{PoS}

\usepackage{cite}

\newcommand{\Figref}[1]{Figure.~\ref{#1}}
\newcommand{\ab}{${\rm ab^{-1}}$~}
\newcommand{\fb}{${\rm fb^{-1}}$~}

\title{Precision Higgs Measurements at the 250 GeV ILC}

\ShortTitle{Precision Higgs Measurements at the 250 GeV ILC}

\author{\speaker{Tomohisa Ogawa}\\
        The Graduate University for Advanced Studies (SOKENDAI) and High Energy Accelerator Research Organization (KEK) \\
        E-mail: \email{ogawat@post.kek.jp}\\
        \\
        \\
        \\
        ~~~~on behalf of Linear Collider Collaboration (LCC) Physics Working Group\\
        \\
	}

\abstract{The plan for the International Linear Collider (ILC) is now being prepared as a staged design, with the first stage of 2~\ab at 250~GeV and later stages achieving the full project specifications with 4~\ab at 500~GeV. The talk presents the capabilities for precision Higgs boson measurements at 250~GeV. It is shown that the 250~GeV stage of the ILC will already provide many compelling results in Higgs physics, with new measurements unavailable at the Large Hadron Collider, model-independent determinations of key parameters, and possible discrimination of a variety of scenarios for new physics.
\\
\\
The talk is based on Phys. Rev. D97 (2018) 053003 and arXiv:1710.07621.
}

\FullConference{39th International Conference on High Energy Physics\\
		4-11 July 2018\\
		Seoul, Korea}

\begin{document}

\section{Introduction}
\vspace{-3mm}
%

The 125~GeV Higgs boson was discovered at the Large Hadron Collider (LHC) in 2012~\cite{1748-0221-3-08-S08001, Higgs1, Higgs2}. Measurements on Higgs couplings and properties with LHC data of approximately 100~\fb, which yield precisions in the 10\%--20\% range, indicate that the new particle is consistent with the Higgs boson in the  Standard Model (SM). Meanwhile, the 13 TeV LHC run has not provided any indications of a new particle. This indicates that additional Higgs bosons or supersymmetric particles, which appear in theories beyond the SM (BSM) such as minimal supersymmetric SM (MSSM) where the Higgs sector is extended as various two Higgs doublet models (2HDMs), will be considerably heavy. 


In such a case, the theories such as the 2HDMs phenomenologically indicate implications that coupling strengths of the lightest BSM Higgs boson to the SM particles will be identical to those of the SM Higgs boson due to suppression depending on the scale of the heaver Higgs boson, which is known as the decoupling limit of the Higgs models~\cite{Haber:1995be}. When generally adapting the decoupling limit for the BSM models in the same manner, typical deviations of the Higgs couplings from the SM expectations due to corrections by the theoretical parameters are predicted to be at a 10\% level or rather smaller~\cite{Baer:2013cma}. Therefore, to detect indications of the deviations and its pattern, reaching precision of 1\% level is required, and the International Linear Collider (ILC) with the 250~GeV operation is expected to be capable of measuring the Higgs couplings with the 1\% precision.  


\vspace{-3mm}
\section{A framework based on effective field theory}
\vspace{-3mm}
Traditionally the $\kappa$-framework has been often employed to evaluate the Higgs couplings~\cite{LHCHiggsCrossSectionWorkingGroup:2012nn}. An application of the $\kappa$-framework, however, is limited only for the leading order consideration because of problems on internal consistency and model-independence confronting the evaluation: 
\vspace{-1mm}
\begin{itemize}
\setlength{\itemsep}{-0.5mm} 
\item There is no relation between the scaling parameters of $\kappa_V$ although symmetry breaking is imposed with the assumption of $SU(2)\times U(1)$ gauge symmetry.     
\item All contributions due to higher-order radiative corrections by the SM particles are regarded as one single $\kappa$ factor although it is composed of several of the other $\kappa$ factors.    
\item Assuming new (next to the leading order) tensor structures composed of field strength tensors, which have momentum dependence, certain production processes and the corresponding decay processes do not match each other in terms of the scaling.
\end{itemize}
\vspace{-1mm}

In contrast, a framework of the Effective Field Theory (EFT) can deal with those problems naturally by adding proper physics-driven higher order field operators. The effective Lagrangian under the ILC-EFT framework has been established based on dimension-6 field operators, where the number of parameters which must be decided simultaneously is 23~\cite{Barklow:2017awn}: the EFT coefficients, renormalizing SM parameters, and Higgs decays unpredicted in the SM.




Advantages of the EFT are not only to deal in the problems but be possible to increase sensitive observables which can contribute to improvement of the Higgs couplings since the EFT can make the parameters relate. Even observables which are not directly related to the Higgs boson become useful to increase the precision. Furthermore, by exploiting beam polarization of the ILC, which is one of the strong functions, the number of sensitive observables can be doubled. Therefore, all the potential of the ILC is possible to be extracted under the EFT framework, within which a percent precision can be reached for the Higgs couplings.



\section{Higgs couplings measurements at 250 GeV}
\vspace{-1mm}
The following observables and information are available to determine the Higgs couplings:
\vspace{-2mm}
\begin{itemize}
\setlength{\itemsep}{-0.5mm} 
\item Direct observations of the Higgs related observables such as inclusive measurement of the $ZZH$ coupling based on the recoil mass technique, various $\sigma \times \rm{BR}$ measurements~\cite{Baer:2013cma}, and also differential cross-section measurements of the Higgs related processes, which are sensitive to new Lorentz structures~\cite{Ogawa:2017bmg}.     
\item Observations on the triple gauge couplings through the $e^+e^- \rightarrow W^+W^-$ process~\cite{Karl:2017let,Marchesini:2011aka}.
\item Electroweak precision observables for constraining the renormalized SM parameters~\cite{Karl:2017let}.
\item Combinations of the beam polarizations can double observables derived from difference of reactions because of the polarization asymmetry.   
\item Higgs observables on $\gamma Z$ and $\gamma \gamma$ provided by the High Luminosity LHC (HL-LHC) operation~\cite{ATL-PHYS-PUB-2014-016}. The ratios of $\rm{BR}_{\gamma Z} / \rm{BR}_{\gamma \gamma}$ and $\rm{BR}_{Z Z} / \rm{BR}_{\gamma \gamma}$ are employed to remove systematics of the measurements.
\end{itemize}
\vspace{-2mm}
\vspace{-2mm}
\begin{figure}[htbp]
	\centering
	\includegraphics[width=110mm]{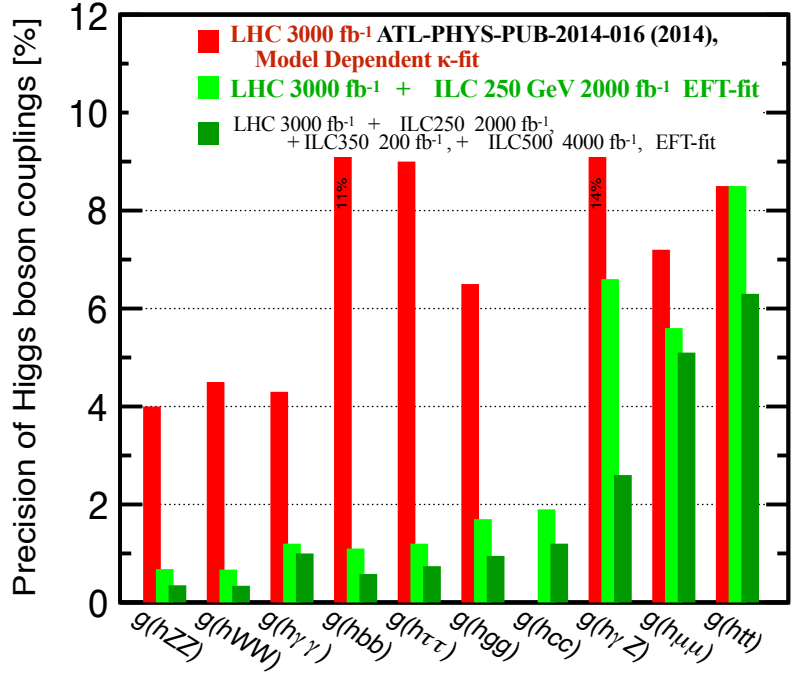}
\caption{The plot is referred from the reference~\cite{Fujii:2017vwa}, which illustrates the precisions of the Higgs couplings evaluated with the global fitting under the ILC-EFT framework (light green). The precisions achievable with ATLAS under the $\kappa$-framework~\cite{ATL-PHYS-PUB-2014-016} is also shown as the comparison (red).}
\label{fig:fig1_1}
\end{figure}

Using these observables mentioned above, global fitting was performed to evaluate the reachable precisions of the Higgs couplings to the SM particles at the 250~GeV ILC, which is described in detail in two papers~\cite{Fujii:2017vwa, Barklow:2017suo}. \Figref{fig:fig1_1} is illustrating the precisions of the Higgs couplings given by ATLAS at HL-LHC, in the $\kappa$-framework and after combination with the 250~GeV ILC under the EFT-framework. The histogram that includes the 250 GeV ILC data indicates that the precisions of the couplings for $ZZ$ and $WW$ reach 1\% below and ones for $b\bar{b}$, $\tau\tau$, and $gg$, which are restricted due to systematics at the LHC, are significantly improved. It is also shown that synergy between HL-LHC and ILC enables improvement in $\gamma\gamma$ and $\gamma Z$ measurements. The precision for $c\bar{c}$ can be achieved with the 250~GeV ILC because of huge QCD backgrounds in the LHC although one for $t\bar{t}$ is not improved due to the given energy. The precisions of the couplings for the vector bosons, up-type and down-type quarks, and third-generation lepton will reach the 1\% level.   


\vspace{-2mm}
\section{Conclusion}
\vspace{-2mm}
We illustrated that the precisions for the Higgs couplings to the SM particles achievable with the 250~GeV ILC is the 1\% level or below, corresponding to the precisions for discriminating the various BSM models that can not be covered by the HL-LHC. The capability of 250 GeV ILC to discriminate the BSM models are discussed in the reference~\cite{Barklow:2017suo} by exemplifying the representative BSM models. The above illustrates that the ILC has a capability to indicate models and a scale of eventual new physics.

\vspace{-2mm}
\bibliographystyle{JHEP} 
\bibliography{skeleton} 


\end{document}